\documentclass[12pt]{article}
\usepackage{epsf,amsmath,euscript}

\def\nslash{\rlap{\hspace{0.02cm}/}{n}}
\def\vslash{\rlap{\hspace{0.02cm}/}{v}}
\def\zslash{\rlap{\hspace{0.02cm}/}{z}}
\def\Aslash{\rlap{\hspace{0.07cm}/}{A}}
\def\calAslash{\rlap{\hspace{0.08cm}/}{{\EuScript A}}}
\def\epsslash{\rlap{\hspace{0.02cm}/}{\varepsilon}}

\def\A{{\EuScript A}}
\def\H{{\EuScript H}}
\def\Q{{\EuScript Q}}
\def\F{{\EuScript F}}
\def\J{{\EuScript J}}

\begin{document}

\begin{titlepage}

\begin{flushright}
CLNS~03/1817\\
SLAC-PUB-9625\\
{\tt hep-ph/0301123}\\[0.2cm]
January 16, 2003
\end{flushright}

\vspace{0.7cm}
\begin{center}
\Large\bf 
Factorization and Sudakov Resummation in\\
Leptonic Radiative B Decay
\end{center}

\vspace{0.8cm}
\begin{center}
{\sc S.W.~Bosch$^a$, R.J.~Hill$^b$, B.O.~Lange$^a$ and M.~Neubert$^a$}\\
\vspace{0.7cm}
{\sl ${}^a$Newman Laboratory for Elementary-Particle Physics, Cornell 
University\\
Ithaca, NY 14853, U.S.A.}\\
\vspace{0.3cm}
{\sl ${}^b$Stanford Linear Accelerator Center, Stanford University\\
Stanford, CA 94309, U.S.A.}
\end{center}

\vspace{1.0cm}
\begin{abstract}
\vspace{0.2cm}\noindent
Soft-collinear effective theory is used to prove factorization of the 
$B\to\gamma l\nu$ decay amplitude at leading power in $\Lambda/m_b$, 
including a demonstration of the absence of non-valence Fock states and 
of the finiteness of the convolution integral in the factorization 
formula. Large logarithms entering the hard-scattering kernel are 
resummed by performing a two-step perturbative matching onto the 
low-energy effective theory, and by solving evolution equations derived 
from the renormalization properties of the leading-order $B$-meson 
light-cone distribution amplitude. As a byproduct, the evolution 
equation for heavy-collinear current operators in soft-collinear 
effective theory is derived.
\end{abstract}
\vfil

\end{titlepage}

\section{Introduction}

The proposal of a ``soft-collinear effective theory'' (SCET) for the 
strong interactions of collinear and soft particles has been an important 
step toward understanding the factorization properties of hard exclusive 
processes in QCD 
\cite{Bauer:2000ew,Bauer:2000yr,Bauer:2001ct,Beneke:2002ph}. In 
particular, it raises the prospects for rigorously proving QCD 
factorization theorems for hadronic and radiative $B$-meson decays into 
light particles such as $B\to\pi\pi$ \cite{Beneke:1999br} and 
$B\to K^*\gamma$ \cite{Beneke:2001at,Bosch:2001gv}, which are of great 
importance to the physics program at the $B$ factories. A challenge 
common to these decays and many others is to understand the interactions 
of collinear particles with the soft spectator quark inside the $B$ 
meson, which give rise to convolutions of hard-scattering kernels with 
$B$-meson light-cone distribution amplitudes (LCDAs). The first 
systematic analysis of these interactions in the framework of SCET has 
recently been performed by two of us \cite{Hill:2002vw}.

The radiative, semileptonic decay $B\to\gamma l\nu$ provides a clean 
environment for the study of soft-collinear interactions 
\cite{Korchemsky:1999qb}. This process is particularly simple in that no 
hadrons appear in the final state. Yet, there is sensitivity to the 
light-cone structure of the $B$ meson, probed by the coupling of the 
high-energy photon to the soft spectator quark inside the heavy meson. In 
the present paper, we apply the formalism of \cite{Hill:2002vw} to prove 
factorization for this decay and systematically resum large Sudakov 
logarithms. The arguments we will present apply, with minimal 
modifications, to more complicated decays such as $B\to K^*\gamma$. 

Several other groups have recently studied the decay $B\to\gamma l\nu$. 
In \cite{Descotes-Genon:2002mw} a QCD factorization formula was 
established at next-to-leading order (NLO) in $\alpha_s$, and it was 
demonstrated that the leading-order LCDA of the $B$ meson is sufficient 
to describe the decay amplitude at leading power in $\Lambda/m_b$ (with 
$\Lambda$ a typical hadronic scale), contrary to the findings of 
\cite{Korchemsky:1999qb}. QCD factorization formulae have also been 
proposed for the related processes $B\to\gamma\gamma$ and 
$B\to\gamma\,l^+ l^-$ \cite{Bosch:2002bv,Descotes-Genon:2002ja}. 
Arguments in favor of factorization in higher orders were given in 
\cite{Lunghi:2002ju} using a formulation of SCET different from the one 
adopted here. 

In the present work we provide the first complete proof of factorization 
for a $B$ decay in which hard-spectator interactions are relevant at 
leading power. We believe that our approach is simpler and more 
transparent than that put forward in \cite{Lunghi:2002ju}. The discussion 
of factorization we will present is more complete in that we prove the 
absence of non-valence Fock-state contributions and the convergence of 
the convolution integral to all orders in perturbation theory. We present
the first correct result for the perturbative hard-scattering kernel in a 
scheme which uses the conventional, mass-independent definition of the
LCDA. In addition, we discuss in detail the complete 
renormalization-group (RG) resummation of large logarithms. To this end, 
we solve evolution equations for the different components of the 
hard-scattering kernel, which follow from the renormalization properties 
of the $B$-meson LCDA. In this context we clarify the connection between 
the anomalous dimension of heavy-collinear currents in SCET and the cusp 
anomalous dimension encountered in the study of Wilson loops with 
light-like segments \cite{Korchemsky:wg,Korchemskaya:1992je}.

Our main goal is to establish the QCD factorization formula 
\cite{Descotes-Genon:2002mw}
\begin{equation}\label{ff}
   {\cal A}(B^-\to\gamma\,l^-\bar\nu_l) 
   \propto m_B f_B\,Q_u \int_0^\infty\!dl_+\,
   \frac{\phi_+^B(l_+,\mu)}{l_+}\,T(l_+,E_\gamma,m_b,\mu)
\end{equation}
to all orders in perturbation theory and at leading power in 
$\Lambda/m_b$. Here $Q_u=\frac23$ is the electric charge of the up-quark
(in units of $e$), $f_B$ is the $B$-meson decay constant, $\phi_+^B$ is 
a leading-order LCDA of the $B$-meson, and $T=1+O(\alpha_s)$ is a 
perturbative hard-scattering kernel. Factorization holds as long as the 
photon is energetic in the $B$-meson rest frame, meaning that $E_\gamma$ 
is of the order of the $b$-quark mass. The physics underlying the 
factorization formula is that a high-energy photon coupling to the soft 
constituents of the $B$-meson produces quantum fluctuations far off their 
mass shell, which can be integrated out in a low-energy effective theory. 
Specifically, when the photon couples to the $b$-quark it takes it off 
shell by an amount of order $m_b^2$, producing a hard quantum fluctuation 
that can be treated using the methods of heavy-quark effective theory 
(HQET) \cite{Neubert:1993mb}. (The resulting contribution to the 
amplitude is, in fact, power suppressed.) When the photon couples to a 
soft light parton inside the $B$ meson it produces a ``hard-collinear'' 
mode that is off shell by an amount of order $m_b\Lambda$. Once these 
short-distance modes are integrated out, the decay amplitude factorizes 
into a soft component (the LCDA) and a hard-scattering kernel.

For the analysis of the $B\to\gamma l\nu$ decay amplitude we work in the 
$B$-meson rest frame and choose the photon momentum along the 
$z$ direction, such that $q^\mu=E_\gamma\,n^\mu$, where $n^\mu=(1,0,0,1)$ 
is a light-like vector. The two transverse polarization states of the 
photon can be expressed in terms of the basis vectors 
$\varepsilon_\mp^\mu=\frac{1}{\sqrt2}(0,1,\mp i,0)$, which correspond to 
left- and right-circular polarization, respectively. It is convenient to 
define a second light-cone vector $\bar n^\mu=(1,0,0,-1)$. Any 4-vector 
can then be expanded as 
$p^\mu=\frac12(p_+\bar n^\mu+p_- n^\mu)+p_\perp^\mu$, where 
$p_+=n\cdot p$ and $p_-=\bar n\cdot p$. 

In order to prove the factorization formula (\ref{ff}) one needs to show 
that \cite{Beneke:2000ry}:
\begin{enumerate}
\item
The decay amplitude can be expanded in powers of transverse momenta and, 
at leading order, can be expressed in terms of a convolution with the 
$B$-meson LCDA as shown in (\ref{ff}). 
\item
After subtraction of infrared contributions corresponding to the 
$B$-meson decay constant and LCDA, the leading contributions to the 
amplitude come from hard internal lines, i.e., the hard-scattering kernel 
$T$ is free of infrared singularities to all orders in perturbation 
theory.
\item
The convolution integral of the hard-scattering kernel with the LCDA is 
convergent.
\item
Non-valence Fock states do not give rise to leading contributions.
\end{enumerate}
In \cite{Lunghi:2002ju} the authors have discussed factorization for the 
decay $B\to\gamma l\nu$, albeit without addressing the last two points in 
this list. (The same criticism applies to the treatment of factorization 
for the decay $B\to D\pi$ presented in \cite{Bauer:2001cu}.) The 
verification of the third point is in essence a check that one has 
correctly identified the infrared degrees of freedom in the effective 
low-energy theory. If the integral in (\ref{ff}) diverged for $l_+\to 0$, 
this would imply that in addition to the hard-collinear modes mentioned 
above one would need to consider long-distance contributions from 
internal momenta collinear with the photon momentum, which are off-shell 
by an amount of order $\Lambda^2$. The fourth point addresses 
contributions from non-valence Fock states, which in the present case 
correspond to $B$-meson matrix elements of trilocal $\bar q\,G^{\mu\nu} b$
 operators. While local operators of this type would be of higher 
dimension and hence power suppressed, non-local operators can contribute 
at leading power if their component fields are smeared over domains of 
size $1/\Lambda$ \cite{Hill:2002vw}. In the following section we will 
address all of the above and, on the example of $B\to\gamma l\nu$ decay, 
provide the first complete proof of a QCD factorization theorem including 
hard-spectator interactions.

Before entering the technical details of the factorization proof, we 
stress that (\ref{ff}) provides the simplest example of a factorization 
formula for a decay in which the hard-scattering mechanism involves three 
different mass scales: a hard scale $E_\gamma\sim m_b$, a soft scale 
$l_+\sim\Lambda$, and an intermediate scale 
$\sqrt{2E_\gamma\,l_+}\sim\sqrt{m_b\Lambda}$. Indeed, a diagrammatic 
analysis of one-loop diagrams for the decay amplitude reveals that 
leading contributions arise from three different regions of loop momenta: 
hard momenta $k\sim m_b$, soft momenta $k\sim\Lambda$, and hard-collinear 
momenta scaling like 
$(k_+,k_-,k_\perp)\sim(\Lambda,m_b,\sqrt{m_b\Lambda})$. This suggests a 
second stage of ``perturbative'' factorization 
\cite{Hill:2002vw,Lunghi:2002ju,Bauer:2002aj}, which we will establish 
below. It says that the hard-scattering kernel itself can be factorized 
as
\begin{equation}\label{fact2}
   T(l_+,E_\gamma,m_b,\mu)
   = H\left(\frac{2E_\gamma}{\mu},\frac{2E_\gamma}{m_b}\right)
   \cdot J\left(\frac{2E_\gamma\,l_+}{\mu^2}\right) .
\end{equation}
The hard component $H$ accounts for the short-distance corrections from 
quantum fluctuations that are off shell by an amount of order $m_b^2$. 
They arise from the coupling of hard or hard-collinear particles to the 
heavy quark. The jet function $J$ accounts for short-distance 
fluctuations that are off shell by an amount of order $m_b\Lambda$, which 
result from the coupling of hard-collinear particles to the light 
spectator quark. The factorization formula (\ref{fact2}) thus separates 
the physics on two different short-distance scales. While this is 
necessary to gain full control over large logarithms arising in the 
perturbative calculation of the hard-scattering kernel, it is not a 
necessary step in the proof of the QCD factorization formula (\ref{ff}), 
which describes the separation of short- and long-distance physics.

\section{Proof of factorization}

We use the formulation of SCET developed in \cite{Hill:2002vw}, where it
was argued that the discussion of exclusive $B$ decays into light 
particles can be made most transparent by matching full QCD amplitudes 
onto a low-energy effective theory in which only long-distance modes are 
kept as dynamical degrees of freedom. In the present case the $B$ meson 
is the only hadron in the process, and so the relevant degrees of freedom 
in the low-energy effective theory are soft partons. The 
strong-interaction Lagrangian consists of the ordinary QCD Lagrangian for 
light quarks and gluons (restricted to the subspace of soft Fourier 
modes) and of the HQET Lagrangian for heavy quarks. In the full theory 
the hadronic part of the decay amplitude is given by the $B$-meson matrix 
element of the time-ordered product of a weak, flavor-changing current 
and the electromagnetic current. At leading order in the effective theory 
this object is matched onto flavor-changing bilocal operators of the form 
$A^{({\rm em})}(z)\,\bar q_s(z)\dots h(0)$, where $A^{({\rm em})}$ is the 
photon field, $h$ is the effective-theory field for the heavy $b$-quark, 
and $q_s$ is the field for a soft $u$-quark. The separation $z$ between 
the fields is nearly light-like, $z^2\sim 1/(m_b\Lambda)\approx 0$. A 
space-time picture of the decay process is as follows: The virtual light 
quark produced at the weak vertex is off shell by an amount of order 
$m_b\Lambda$ and almost collinear with the photon direction. The 
electromagnetic vertex is thus located at a distance 
$z^\mu=tn^\mu+s\bar n^\mu+z_\perp^\mu$ relative to the weak vertex, where 
$t\sim 1/\Lambda$ is the parametrically largest component. Because of the 
scaling properties of the photon and soft spectator momenta one can 
replace $\bar q_s(z)\simeq\bar q_s(tn)$ and 
$A^{({\rm em})}(z)\simeq A^{({\rm em})}(s\bar n)$ to leading power. From 
the discussion in \cite{Hill:2002vw}, it follows that at leading order in 
$\Lambda/m_b$ there exists a unique type of SCET operators that mediate 
this decay and are allowed by gauge and reparameterization invariance. 
They are
\begin{eqnarray}\label{ops}
   &&\sum_{q=u,b} ie\,Q_q \int d^4x\,T\,\Big\{
    \big[ \bar u\gamma^\mu(1-\gamma_5) b \big](0),
    \big[ \bar q\,\Aslash^{\rm (em)}\,q \big](x) \Big\} \nonumber\\
   &\to& \sum_i \int ds\,dt\,\widehat C_i(t,s,v\cdot q,m_b,\mu)\,
    \bar\Q_s(tn)\,\calAslash_{c\perp}^{({\rm em})}(s\bar n)\,
    \frac{\nslash}{2}\,\Gamma_i\,\H(0) \nonumber\\
   &=& \sum_i \int dt\,\widetilde C_i(t,\bar n\cdot q,v\cdot q,m_b,\mu)\,
    \bar q_s(tn)\,S(tn,0)\,\calAslash_{c\perp}^{({\rm em})}(0)\,
    \frac{\nslash}{2}\,\Gamma_i\,h(0) \,,
\end{eqnarray}
where $\H=S^\dagger h$ and $\Q_s=S^\dagger q_s$ with the path-ordered
exponential 
\begin{equation}
   S(x) = P\exp\left( ig\int_{-\infty}^0\!dw\,n\cdot A_s(x+wn) \right)
\end{equation}
are gauge-invariant combinations of SCET fields and Wilson lines, and 
$A_s$ is the soft gluon field. The combination 
$S(x,y)\equiv S(x)\,S^\dagger(y)$ appearing in the last line of 
(\ref{ops}) represents a soft Wilson line connecting the two points $x$ 
and $y$ on a straight segment. Physically, this string operator arises 
because the virtual quark propagating between the two vertices can emit 
multiple soft gluons without power suppression. The quantity 
$\calAslash_{c\perp}^{({\rm em})}$ is the electromagnetic analog of the 
gauge-invariant collinear gluon field defined in \cite{Hill:2002vw}. To 
first order in $e$ we have
\begin{equation}
   \calAslash_{c\perp}^{({\rm em})}(0)
   = \bar n_\alpha\gamma_\mu^\perp
   \int_{-\infty}^0\!dw\,e F^{\alpha\mu}(w\bar n) \,.
\end{equation}
The Feynman rule for this object is simply $e\,\epsslash^*$, where 
$\varepsilon$ is the photon polarization vector. Finally, from the fact 
that the leptonic weak current $\bar\nu\gamma_\mu(1-\gamma_5)\,l$ is 
conserved (in the limit where the lepton mass is neglected) it follows 
that the relevant Dirac structures $\Gamma_i$ in (\ref{ops}) can be taken 
as $\Gamma_1=\gamma^\mu(1-\gamma_5)$ and $\Gamma_2=n^\mu(1+\gamma_5)$.

In the last step in (\ref{ops}) we have used that the photon momentum $q$ 
is an external momentum, so that the integration over $s$ can be 
performed and leads to new coefficients 
$\widetilde C_i(t,\bar n\cdot q,v\cdot q,m_b,\mu)
=\int ds\,e^{is\bar n\cdot q}\,\widehat C_i(t,s,v\cdot q,m_b,\mu)$. Note 
that the scalar products $\bar n\cdot q=2E_\gamma$ and 
$v\cdot q=E_\gamma$ are both determined in terms of the photon energy; 
however, for a while it will be useful to distinguish between these two 
variables.

SCET power counting shows that the soft fields $\H$ and $\Q_s$ scale like 
$\Lambda^{3/2}$, the integration variable $t$ scales like $1/\Lambda$, 
and the Wilson coefficients $\widetilde C_i$ scale like 1. It follows 
that the hadronic components of the SCET operators on the right-hand side 
of (\ref{ops}) scale like $\Lambda^2$, which is one power of $\Lambda$ 
less than a local current containing two soft quark fields. That one can 
write down such operators is a consequence of the fact that transverse 
collinear fields have unsuppressed interactions with soft light quarks 
\cite{Hill:2002vw}. As a result, at leading power the Wilson coefficients 
$\widetilde C_i$ receive contributions only from Feynman diagrams with a 
photon attached to the light spectator quark in the $B$ meson.

Following \cite{Grozin:1996pq}, we define the two leading-order LCDAs for 
the $B$ meson in position space in terms of the HQET matrix element
\begin{eqnarray}\label{LCDA}
   &&\frac{1}{\sqrt{m_B}}\,
    \langle\,0\,|\,\bar q_s(z)\,S(z,0)\,\Gamma\,h(0)\,|\bar B(v)\rangle
    \nonumber\\
   &=& - \frac{iF(\mu)}{2}\,
    \mbox{tr}\bigg[ \left( \widetilde\phi_+^B(\tau,\mu)
    - \frac{\zslash}{2\tau}\,
    \Big[ \widetilde\phi_-^B(\tau,\mu) - \widetilde\phi_+^B(\tau,\mu)
    \Big] \right) \Gamma\,\frac{1+\vslash}{2}\,\gamma_5 \bigg] \,.
\end{eqnarray}
Here $z^2=0$ is a null vector, $v$ is the $B$-meson velocity, and 
$\tau=v\cdot z-i0$. The quantity $F(\mu)$ corresponds to the asymptotic 
value of the product $f_B\sqrt{m_B}$ in the heavy-quark limit 
\cite{Neubert:1991sp}. The two distribution amplitudes obey the 
normalization $\widetilde\phi_\pm^B(0,\mu)=1$ at $\tau=0$. In our case 
$z=tn$, and because of the presence of the factor $\nslash$ in 
(\ref{ops}) it follows that only the function 
$\widetilde\phi_+^B(\tau,\mu)$ contributes, to which we refer as 
{\em the\/} leading-order LCDA. Note that with our choice of the 
light-cone basis vector $n$ we have $\tau=t\,v\cdot n=t$. Performing the 
relevant traces over Dirac matrices, we find that at leading power in 
$\Lambda/m_b$ the decay amplitude vanishes if the photon has 
right-circular polarization, while for a photon with left-circular 
polarization it is given by
\begin{eqnarray}\label{ampl}
   {\cal A}(B^-\to\gamma_L\,l^-\bar\nu_l) 
   &=& \frac{iG_F}{\sqrt2}\,V_{ub}\,e\,
    \bar u_l(p_l)\,\epsslash_-^*(1-\gamma_5) v_\nu(p_\nu) \nonumber\\
   &\times& \sqrt{m_B}\,F(\mu) \int dt\,
    \widetilde C_1(t,\bar n\cdot q,v\cdot q,m_b,\mu)\,
    \widetilde\phi_+^B(t,\mu) + \dots \,,
\end{eqnarray}
where the dots represent power-suppressed contributions. Only the SCET 
operator with Dirac structure $\Gamma_1=\gamma^\mu(1-\gamma_5)$ 
contributes to the decay amplitude. In order to recast the above result 
in a form resembling the factorization formula (\ref{ff}) we introduce 
the Fourier transforms of the Wilson coefficient function and the LCDA as 
\cite{Hill:2002vw,Grozin:1996pq}
\begin{equation}\label{FT}
\begin{aligned}
   C_1(l_+,\bar n\cdot q,v\cdot q,m_b,\mu)
   &= \int dt\,e^{-il_+ t}\,
    \widetilde C_1(t,\bar n\cdot q,v\cdot q,m_b,\mu) \,, \\
   \phi_+^B(\omega,\mu)
   &= \frac{1}{2\pi} \int dt\,e^{i\omega t}\,\widetilde\phi_+^B(t,\mu)
    \,.
\end{aligned}
\end{equation}
The analytic properties of the function $\widetilde\phi_+^B(t,\mu)$ in 
the complex $t$-plane imply that $\phi_+^B(\omega,\mu)=0$ if $\omega<0$.
Finally, the HQET parameter $F(\mu)$ is related to the physical $B$-meson 
decay constant through 
$f_B\sqrt{m_B}=K_F(m_b,\mu)\,F(\mu)\,[1+O(\Lambda/m_b)]$, where at NLO in 
the $\overline{\rm MS}$ scheme \cite{Neubert:1991sp}
\begin{equation}
   K_F(m_b,\mu) = 1 + \frac{C_F\,\alpha_s(\mu)}{4\pi}
   \left( 3\ln\frac{m_b}{\mu} - 2 \right) .
\end{equation}
Combining these results, it follows that the terms shown in the second 
line of (\ref{ampl}) equal those on the right-hand side of the 
factorization formula (\ref{ff}) if we identify the hard-scattering 
kernel as  
\begin{equation}\label{kernel}
   \frac{Q_u}{l_+}\,T(l_+,E_\gamma,m_b,\mu)
   = K_F^{-1}(m_b,\mu)\,C_1(l_+,2E_\gamma,E_\gamma,m_b,\mu) \,.
\end{equation}

Let us now discuss factorization in the context of our formalism. The 
fact that the position-space SCET operators appearing on the right-hand 
side of (\ref{ops}) contain component fields with light-like separation 
implies that transverse parton momenta can be set to zero at leading 
power. As we have seen, this naturally leads to the appearance of LCDAs. 
In SCET the hard-scattering kernel $T$ is identified with a Wilson 
coefficient. The absence of infrared singularities then follows from the 
very existence of a low-energy effective theory, because Wilson 
coefficients arise from matching and by construction are insensitive to 
infrared physics. At this point we have achieved as much as 
\cite{Lunghi:2002ju}.

We proceed to prove the convergence of the convolution integral in 
(\ref{ff}). The key ingredient here is to note that the invariance of 
SCET operators under reparameterizations of the light-cone basis vectors 
$n$ and $\bar n$ \cite{Manohar:2002fd} can be used to deduce the 
dependence of Wilson coefficient functions on the separation $t$ between 
the component fields of non-local operators \cite{Hill:2002vw}. In our 
case, invariance of the operators in (\ref{ops}) under the rescaling
transformation $n^\mu\to n^\mu/\alpha$ and 
$\bar n^\mu\to\alpha\bar n^\mu$ (with fixed $v$) implies that
\begin{eqnarray}\label{HiJdef}
   \widetilde C_i(t,\bar n\cdot q,v\cdot q,m_b,\mu)
   &=& \widetilde C_i(\alpha t,\alpha\bar n\cdot q,v\cdot q,m_b,\mu)
    \nonumber\\
   &=& \widetilde H_i(v\cdot q,m_b,\mu)
    \cdot\widetilde J(\alpha t,\alpha\bar n\cdot q,\mu)
\end{eqnarray}
to all orders in perturbation theory. In other words, the variables $t$ 
and $\bar n\cdot q$ can only appear in the combination $\bar n\cdot q/t$, 
but not individually. (Similarly, the variable $t$ enters the argument of 
the LCDA in (\ref{LCDA}) in the reparameterization-invariant combination 
$\tau=t\,v\cdot n$.) In the second step we have used the fact that 
$v\cdot q$ and $m_b$ enter only through interactions of hard or 
hard-collinear gluons with the heavy quark. The corresponding modes can 
be integrated out in a first matching step and lead to the functions 
$\widetilde H_i$, which depend on the Dirac structure of the weak 
current containing the heavy quark. The non-localities of the component 
fields in (\ref{ops}) result from the coupling of hard-collinear fields 
to the soft spectator quark in the $B$ meson. These effects live on 
scales of order $m_b\Lambda$ and can be integrated out in a second step, 
leading to the function $\widetilde J$. Since the coefficients 
$\widetilde C_i$ are dimensionless, it follows that (with a slight abuse
of notation)
\begin{equation}\label{Ciscal}
   \widetilde C_i(t,\bar n\cdot q,v\cdot q,m_b,\mu)
   = \widetilde H_i\left(\frac{2v\cdot q}{\mu},x_\gamma\right)
   \cdot\widetilde J\left(\frac{\bar n\cdot q}{\mu^2\,t}\right) ,
\end{equation}
where $x_\gamma\equiv 2v\cdot q/m_b=2E_\gamma/m_b$ is a scaling variable 
of order 1. Corrections to this perturbative factorization formula are 
suppressed by a ratio of the intermediate scale 
$\bar n\cdot q/t\sim m_b\Lambda$ and a hard scale of order $m_b^2$. 
(While the fact that these corrections scale like $\Lambda/m_b$ is an 
immediate consequence of our discussion here, it is not obvious in the 
context of the approach proposed in \cite{Lunghi:2002ju,Bauer:2002aj}, 
which is based on an expansion in powers of $\sqrt{\Lambda/m_b}$.) The 
corresponding result for the hard-scattering kernel obtained after 
Fourier transformation has the form shown in (\ref{fact2}) if we identify
\begin{equation}\label{ids}
\begin{aligned}
   H\left(\frac{2v\cdot q}{\mu},x_\gamma\right)
   &= K_F^{-1}(m_b,\mu)\,
    \widetilde H_1\left(\frac{2v\cdot q}{\mu},x_\gamma\right) , \\
   \frac{Q_u}{l_+}\,J\left(\frac{\bar n\cdot q\,l_+}{\mu^2}\right)
   &= \int dt\,e^{-il_+ t}\,
    \widetilde J\left(\frac{\bar n\cdot q}{\mu^2\,t}\right) ,
\end{aligned}
\end{equation}
where $\bar n\cdot q\,l_+=2q\cdot l=2E_\gamma\,l_+$. Since the dependence
of the coefficient functions on the renormalization scale is logarithmic, 
it follows that to all orders in perturbation theory the Wilson 
coefficients in (\ref{Ciscal}) scale like $\widetilde C_i\sim 1$ modulo 
logarithms. (Correspondingly, the kernel scales like $T\sim 1$ modulo 
logarithms.) The convergence of the convolution integral in (\ref{ampl})
in the infrared region $t\to\infty$, corresponding to the region 
$l_+\to 0$ in the factorization formula (\ref{ff}), then follows to all 
orders in perturbation theory as long as the integral converges at tree 
level. Because the $B$ meson has a spatial size of order $1/\Lambda$ due 
to confinement, the bilocal matrix element must vanish faster than $1/t$ 
for $t\gg 1/\Lambda$, and so the integral over $t$ is convergent.

The absence of endpoint divergences in convolution integrals is tied to 
the question of whether the infrared degrees of freedom have correctly 
been identified in the effective theory \cite{Bauer:2002aj}. Let us 
illustrate this point on the case at hand. Assume the integral over $l_+$ 
in (\ref{ff}) diverged logarithmically for $l_+\to 0$. This would 
indicate that there were leading-power contributions to the decay 
amplitude from regions where intermediate propagators (such as the quark 
propagator in Figure~\ref{fig:tree} below) are collinear with the 
external photon. In addition to the modes we call hard-collinear one then 
had to include other, long-distance collinear modes in the construction 
of the effective theory, which would invalidate factorization. The decay 
amplitude would then be sensitive to the hadronic structure of the photon 
as well as the full Bethe--Salpeter wave function of the $B$ meson. Our 
demonstration that the region $l_+\to 0$ is power suppressed is thus an 
essential ingredient of the factorization proof.

The final step in the factorization proof is to demonstrate the power 
suppression of more complicated projections onto the $B$ meson involving 
higher Fock states or transverse parton momenta. The fact that such 
projections do not contribute at leading power follows from the rules for 
constructing SCET operators out of gauge-invariant building blocks, as 
explained in \cite{Hill:2002vw}. Projections sensitive to transverse 
momentum components contain extra derivatives and so are power 
suppressed. Projections corresponding to non-valence Fock states contain 
insertions of the soft (subscript ``$s$'') gluon field
\begin{equation}
   \A_s^\nu(x) = \big[ S^\dagger(iD_s^\mu S) \big](x)
   = \int_{-\infty}^0\!dw\,n_\alpha\,\big[ S^\dagger g G_s^{\alpha\nu}
   S \big](x+wn) \,.
\end{equation}
Since $\A_s$ scales like $\Lambda$, such insertions lead to power 
suppression unless this field is integrated over a domain of extension 
$1/\Lambda$. It follows that the only possibility for a leading-power
contribution from non-valence Fock states would be to include a factor of 
$\int du\,\A_s^\nu(un)$ with $u\sim 1/\Lambda$ (up to dimensionless 
Wilson coefficients of order 1). An example of such an HQET operator 
(omitting gauge strings and Dirac structures) is
$\int dt\int dw\,w\,\bar q_s(tn)\,n_\alpha G_s^{\alpha\nu}(wn)\,h(0)$, 
which is indeed of leading order $\sim\Lambda^2$ in power counting.
Reparameterization invariance requires that this object must be 
accompanied by an additional factor of $n$ in the numerator.\footnote{The 
reason is that the integration variable $u$ scales like $\bar n$, and 
that the basis vectors $n$ and $\bar n$ always come in pairs since they 
are introduced through the light-cone decomposition of 4-vectors. The 
only alternative to having a factor of $n$ in the numerator would be to 
add a factor of $\bar n/(t\,\bar n\cdot q)$ or 
$\bar n/(u\,\bar n\cdot q)$, both of which scale like $\Lambda/m_b$ and 
thus would lead to power suppression.}
Since $n\cdot\A_s=0$ by definition, the only possibility would be to 
include an insertion of $\int du\,\nslash\,\calAslash_{s\perp}(un)$ 
somewhere between the light-quark field $\bar\Q_s$ and the Dirac matrix 
$\Gamma_i$ in the SCET operators in (\ref{ops}). However, any such 
insertion vanishes, since $n^2=0$. It is important for this argument that 
the heavy quark can be integrated out before one removes the off-shell 
modes resulting from soft-collinear interactions. The heavy quark 
therefore decouples from such interactions. This ensures that the factor 
$\nslash$ cannot appear to the right of $\Gamma_i$, and it excludes the 
appearance of $v\cdot n$ instead of $\nslash$. 

We have thus completed the proof of the factorization formula (\ref{ff}) 
to all orders in perturbation theory, and at leading power in 
$\Lambda/m_b$. The remainder of this paper is devoted to the calculation 
of the hard-scattering kernel at NLO in RG-improved perturbation theory, 
including a complete resummation of large logarithms.

\section{Calculation of the hard-scattering kernel}

\begin{figure}
\epsfxsize=14.0cm
\centerline{\epsffile{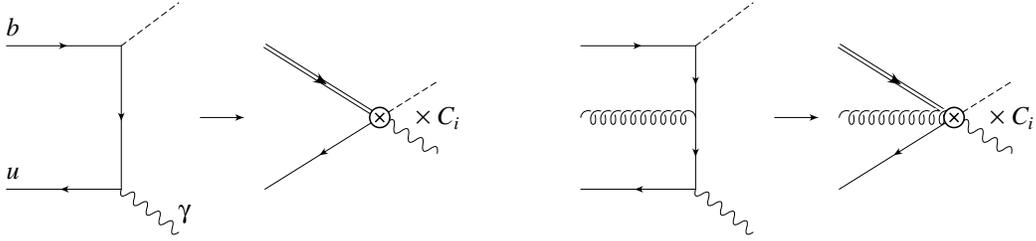}}
\vspace{0.0cm}
\centerline{\parbox{14cm}{\caption{\label{fig:tree}
Tree-level matching calculation for the Wilson coefficients $C_1$ and 
$C_2$, without (left) and with (right) an external soft gluon. The dashed 
line denotes the flavor-changing weak current. The resulting non-local 
operators in SCET are denoted by a crossed circle. Double lines represent 
effective heavy-quark fields in HQET.}}}
\end{figure}

The Wilson coefficients $\widetilde C_i$ in (\ref{ops}) are derived by 
matching perturbative expressions for operator matrix elements in the 
full theory onto corresponding expressions in the effective theory. 
Because by construction the Wilson coefficients are insensitive to 
infrared physics the matching can be done using on-shell external quark 
states. We thus assign incoming momenta $m_b\,v$ to the heavy quark and 
$l$ (with $l^2=0$) to the soft light quark. At tree-level the relevant 
amplitude in the full theory is obtained from the first Feynman diagram 
shown in Figure~\ref{fig:tree}. At leading power only photon emission 
from the light spectator quark contributes; emission from the $b$-quark 
is suppressed by one power of $\Lambda/m_b$. The corresponding amplitude 
in the effective theory follows from the second diagram. A 
straightforward matching calculation yields for the Wilson coefficients 
at this order $C_1=Q_u/l_+$ and $C_2=0$, where $l_+=n\cdot l-i0$. (In 
general, the variable $l_+$ is conjugate to the coordinate $t$ in 
(\ref{ops}). In the present case, $l_+$ coincides with the plus component 
of the spectator momentum because of the particular external state we 
choose for the matching calculation.) The corresponding results in 
position space are $\widetilde C_1=iQ_u\,\theta(t)$ and 
$\widetilde C_2=0$. 

While these results are most easily derived by matching amplitudes with
two external quarks, the Wilson coefficients are independent of the 
nature of the external states. Alternatively, therefore, they can be
determined by matching amplitudes with external soft gluons. Gluon 
emissions from the external quark lines cancel in the matching, since 
those emissions are the same in the full theory and in SCET. However, 
gluon emissions from the internal quark propagator, which is integrated 
out in the effective theory, are contained in the Wilson line $S(tn,0)$, 
which sums up an infinite number of soft gluon insertions. The one-gluon 
example is illustrated in the last two diagrams in Figure~\ref{fig:tree}. 
Evaluating these graphs one readily recovers the results given above. In 
that way one confirms (at tree level) our general result about the 
absence of operators containing additional insertions of the soft gluon 
field $\A_s$.

\begin{figure}
\epsfxsize=14.0cm
\centerline{\epsffile{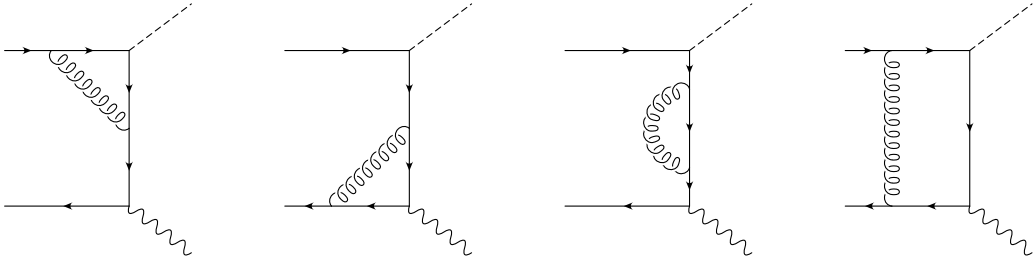}}
\vspace{0.0cm}
\centerline{\parbox{14cm}{\caption{\label{fig:QCD}
One-loop diagrams in the full theory contributing at leading power to the 
$B\to\gamma l\nu$ decay amplitude.}}}
\end{figure}

Beyond tree level the coefficient functions can be written in the form
\begin{equation}
   C_i = \frac{Q_u}{l_+} \left[ \delta_{i1}
   + \frac{C_F\,\alpha_s(\mu)}{4\pi}\,c_i + \dots \right] .
\end{equation}
To obtain the NLO corrections $c_i$ we evaluate the one-loop 
contributions to the decay amplitude in the full theory and in SCET. 
The relevant diagrams in full QCD are shown in Figure~\ref{fig:QCD}. In 
addition there is a contribution from the wave-function renormalization 
for the heavy quark. Using the $\overline{\rm MS}$ regularization scheme 
with $d=4-2\epsilon$ dimensions and anti-commuting $\gamma_5$, we find 
the following expressions for the contributions $A_i$ corresponding to 
the coefficients $c_i$ (before subtraction of the pole terms):
\begin{equation}\label{AiQCD}
\begin{aligned}
   A_1^{\rm QCD}
   &= \left( \frac{m_b}{\mu} \right)^{-2\epsilon}
    \bigg[ - \frac{1}{2\epsilon} - \ln^2\frac{2E_\gamma\,l_+}{m_b^2}
    - 2(1-2\log x_\gamma)\,\ln\frac{2E_\gamma\,l_+}{m _b^2}
    - 4\ln^2 x_\gamma \\
   &\hspace{2.6cm}\mbox{}+ \frac{2-3x_\gamma}{1-x_\gamma}\,\ln x_\gamma
    - 2L_2(1-x_\gamma) - 2 - \pi^2 \bigg] \\
   &+ \left( \frac{2E_\gamma\,l_+}{\mu^2} \right)^{-\epsilon}
    \left( - \frac{2}{\epsilon} - 5 \right) 
    + \left( \frac{-2v\cdot l}{\mu} \right)^{-2\epsilon}
    \bigg[ - \frac{1}{\epsilon^2}
    + \ln^2\left(\frac{-2v\cdot l}{l_+} \right) 
    + \frac{7\pi^2}{12} \bigg] \,, \\
   A_2^{\rm QCD}
   &= \left( \frac{m_b}{\mu} \right)^{-2\epsilon}\,
    \frac{x_\gamma\ln x_\gamma}{1-x_\gamma} \,.
\end{aligned}
\end{equation}
In the expression for $A_1^{\rm QCD}$ the first bracket contains the 
vertex correction for the weak current and the contribution from 
wave-function renormalization for the heavy quark, the second term 
corresponds to the vertex correction for the electromagnetic current and 
the self-energy insertion on the intermediate quark propagator, and the 
last term corresponds to the box diagram. Note that the various terms 
depend on three different mass scales: $m_b$ (hard), 
$2E_\gamma\,l_+\sim m_b\Lambda$ (hard-collinear), and $l\sim\Lambda$ 
(soft). Whereas the first two scales are perturbative, the contribution 
from the box diagram is dominated by soft physics. Indeed, power counting 
shows that the hard and collinear contributions to the box graph only 
contribute at subleading power in $\Lambda/m_b$, as has been observed 
previously in \cite{Descotes-Genon:2002mw}. Note that the box 
contribution involves components of the soft spectator momentum other 
than $l_+$ \cite{Korchemsky:1999qb}. However, this dependence will cancel 
in the matching. Finally, it is worth emphasizing that there is no 
leading contribution to the decay amplitude from the region of loop 
momenta that are collinear (not hard-collinear) with the external photon 
momentum. If it were present, this would imply that non-perturbative 
effects related to the hadronic structure of the photon would contribute 
at leading power, invalidating the factorization formula (\ref{ff}).

\begin{figure}
\epsfxsize=10.0cm
\centerline{\epsffile{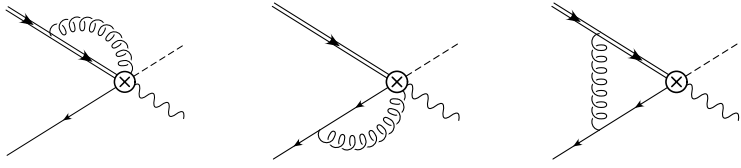}}
\vspace{0.0cm}
\centerline{\parbox{14cm}{\caption{\label{fig:SCET}
One-loop diagrams in the effective theory whose contribution to the 
amplitude needs to be subtracted in the calculation of the Wilson 
coefficients.}}}
\end{figure}

In the next step we evaluate the corresponding contributions at one-loop 
order in the effective theory. Generically, they are of the form
$C_{\rm 1-loop}\,\otimes \langle O_{\rm SCET}\rangle_{\rm tree}
+C_{\rm tree}\,\otimes \langle O_{\rm SCET}\rangle_{\rm 1-loop}$. The 
second contribution, which involves one-loop matrix elements of SCET 
operators convoluted with tree-level coefficient functions, must be 
subtracted from the amplitudes obtained in the full theory. The relevant 
diagrams are shown in Figure~\ref{fig:SCET}. We find
\begin{eqnarray}\label{AiSCET}
   A_1^{\rm SCET}
   &=& \left( \frac{l_+}{\mu} \right)^{-2\epsilon}
    \left( - \frac{1}{\epsilon^2} - \frac{3\pi^2}{4} \right) 
    + \left( \frac{-2v\cdot l}{\mu} \right)^{-2\epsilon}
    \bigg[ - \frac{1}{\epsilon^2}
    + \ln^2\left(\frac{-2v\cdot l}{l_+} \right) 
    + \frac{7\pi^2}{12} \bigg] \,, \nonumber\\
   A_2^{\rm SCET} &=& 0 \,.
\end{eqnarray}
The first term in the expression for $A_1^{\rm SCET}$ corresponds to the 
first two diagrams in the figure, while the second term is obtained from 
the last graph. Note that this is precisely the same contribution as 
obtained from the box diagram in the full theory. 

The difference of the expressions given in (\ref{AiQCD}) and 
(\ref{AiSCET}) determines the NLO contributions to the Wilson coefficient 
functions. Subtracting the pole terms in the $\overline{\rm MS}$ scheme, 
we find
\begin{equation}\label{ci}
\begin{aligned}
   c_1 &= - 2\ln^2\frac{m_b}{\mu} + (5-4\ln x_\gamma) \ln\frac{m_b}{\mu}
    + \ln^2\frac{2E_\gamma\,l_+}{\mu^2} - 2\ln^2 x_\gamma \\
   &\hspace{0.5cm}\mbox{}+ \frac{2-3x_\gamma}{1-x_\gamma}\,\ln x_\gamma
    - 2L_2(1-x_\gamma) - 7 - \frac{\pi^2}{4} \,, \\
   c_2 &= \frac{x_\gamma\ln x_\gamma}{1-x_\gamma} \,.
\end{aligned}
\end{equation}
Our result for the Wilson coefficient $C_1$ agrees with the expressions
for the hard-scattering kernel given in \cite{Descotes-Genon:2002mw} and 
\cite{Lunghi:2002ju}; however, eq.~(\ref{kernel}) shows that identifying 
the kernel with $C_1$ misses the large logarithms encountered when the 
HQET matrix element $F(\mu)$ is related to the physical $B$-meson decay
constant. (In other words, these papers implicitly assume an 
$m_b$-dependent definition and normalization of the $B$-meson LCDA, which 
is unconventional. In our approach the LCDA is normalized to unity, and 
all mass dependence is explicit in the hard-scattering kernel.) Including 
these corrections and simplifying the answer, we obtain for the 
hard-scattering kernel at NLO the final result
\begin{eqnarray}\label{Tres}
   T(l_+,E_\gamma,m_b,\mu)
   &=& 1 + \frac{C_F\,\alpha_s(\mu)}{4\pi}
    \bigg[ - 2\ln^2\frac{2E_\gamma}{\mu} + 2\ln\frac{2E_\gamma}{\mu}
    + \ln^2\frac{2E_\gamma\,l_+}{\mu^2} \nonumber\\
   &&\hspace{2.63cm}\mbox{}- \frac{x_\gamma\ln x_\gamma}{1-x_\gamma}
    - 2L_2(1-x_\gamma) - 5 - \frac{\pi^2}{4} \bigg] \,.
\end{eqnarray}
Note that, despite appearance, the scale dependence of the expression in 
brackets is of the form 
$2\ln^2(\mu/l_+)-2\ln(\mu/l_+)+\mbox{$\mu$-independent terms}$, 
indicating that it can be canceled by the scale dependence of the LCDA 
$\phi_+^B(l_+,\mu)$ under the convolution integral in (\ref{ff}). This 
would be impossible if the argument of the $\ln^2$-term depended on the 
hard scales $E_\gamma$ or $m_b$.

\section{RG evolution and resummation}

The hard-scattering kernel $T$ contains the logarithms 
$\ln(2E_\gamma/\mu)$ and $\ln(2E_\gamma\,l_+/\mu^2)$, which cannot be 
made small simultaneously for any choice of the renormalization scale 
$\mu$. While for realistic values of the photon energy 
($E_\gamma\sim m_b/2\sim 2.5$\,GeV) these logarithms are numerically not 
too large, it is conceptually interesting to gain control over 
the perturbative expansion of the kernel by summing the various 
logarithms to all orders in perturbation theory.

It was proposed in \cite{Descotes-Genon:2002mw} to choose the 
renormalization scale of order $\mu^2\sim m_b\Lambda$, and to identify 
the remaining large logarithms in (\ref{ci}) with those arising in the 
matching of heavy-collinear current operators onto SCET. The precise 
nature of this identification beyond one-loop order was however left 
unclear. In the spirit of an effective-theory approach one would rather 
prefer to take the renormalization scale further down to a value 
$\mu=\mbox{few}\times\Lambda_{\rm QCD}$ independent of the $b$-quark mass 
and the photon energy, yet large enough for a perturbative treatment. In 
this way all dependence on the large scales $m_b$ and $E_\gamma$ becomes 
explicit and is contained in Wilson coefficients of the effective theory.

To gain full control over the large logarithms in the perturbative 
expansion of the kernel we perform the matching onto SCET in two steps 
\cite{Hill:2002vw,Lunghi:2002ju,Bauer:2002aj}. In the first step the 
off-shell fluctuations of the heavy $b$ quark are integrated out by 
matching onto HQET. Hard-collinear modes with momenta scaling like 
$(k_+,k_-,k_\perp)\sim(\Lambda,m_b,\sqrt{m_b\Lambda})$ are integrated out
in a second step. In contrast with \cite{Lunghi:2002ju}, we avoid the
explicit construction of the intermediate effective theory. Since this 
theory is needed only for RG improvement, it suffices to perform the 
matching diagrammatically using the method of regions 
\cite{Smirnov:1996ng,Beneke:1997zp}. This approach allows us to compute 
the functions $H$ and $J$ in (\ref{fact2}) systematically order by order 
in perturbation theory. Specifically, because the intermediate mass scale 
of order $m_b\Lambda$ always arises from a scalar product of a soft 
momentum with a collinear momentum, one can perturbatively match onto the 
intermediate theory by simply setting the soft momentum $l=0$. In that 
way only hard fluctuations associated with couplings to the heavy quark 
are integrated out. This yields precisely the function $H$. The jet 
function then follows from the ratio $J=T/H$.

Let us illustrate this for the case at hand. For $l=0$ the second and 
third diagrams in Figure~\ref{fig:QCD} involve scaleless integrals that 
vanish in dimensional regularization, while the box graph can readily be 
shown to vanish at leading power. The remaining contribution from the 
weak vertex correction and wave-function renormalization for the heavy 
quark yields
\begin{equation}\label{Ail0}
   A_1^{\rm QCD}\Big|_{l=0}
   = \left( \frac{m_b}{\mu} \right)^{-2\epsilon}
   \bigg[ - \frac{1}{\epsilon^2} - \frac{5}{2\epsilon}
   + \frac{2\ln x_\gamma}{\epsilon} - 2\ln^2 x_\gamma
   + \frac{2-3x_\gamma}{1-x_\gamma}\,\ln x_\gamma
   - 2L_2(1-x_\gamma) - 6 - \frac{\pi^2}{12} \bigg] \,,
   \phantom{\Bigg|}
\end{equation}
while the expression for $ A_2^{\rm QCD}\big|_{l=0}$ is the same as that 
for $ A_2^{\rm QCD}$ given in (\ref{AiQCD}). All graphs in the effective
theory involve tadpole integrals and vanish in dimensional 
regularization. Hence, after $\overline{\rm MS}$ subtractions the above 
result determines the hard function $\widetilde H_1$ defined in 
(\ref{HiJdef}). Using the first relation in (\ref{ids}), it then follows 
that 
\begin{equation}\label{Hres}
   H\left(\frac{2E_\gamma}{\mu},x_\gamma\right)
   = 1 + \frac{C_F\,\alpha_s(\mu)}{4\pi}
   \bigg[ - 2\ln^2\frac{2E_\gamma}{\mu} + 2\ln\frac{2E_\gamma}{\mu}
   - \frac{x_\gamma\ln x_\gamma}{1-x_\gamma} - 2L_2(1-x_\gamma)
   - 4 - \frac{\pi^2}{12} \bigg] \,.
   \phantom{\Bigg|}
\end{equation}
According to (\ref{fact2}), the difference between (\ref{Tres}) and 
(\ref{Hres}) determines the one-loop contribution to the jet function, 
which is thus given by
\begin{equation}\label{Jres}
   J\left(\frac{2E_\gamma\,l_+}{\mu^2}\right)
   =  1 + \frac{C_F\,\alpha_s(\mu)}{4\pi}
    \bigg( \ln^2\frac{2E_\gamma\,l_+}{\mu^2} - 1 - \frac{\pi^2}{6}
    \bigg) .
\end{equation}

In order to proceed we need RG equations obeyed by the various 
coefficient functions. The fact that the decay amplitude in (\ref{ff}) 
is scale independent links the scale dependence of the hard-scattering 
kernel to the evolution of the LCDA $\phi_+^B(\omega,\mu)$ 
\cite{Grozin:1996pq}. By analyzing the renormalization properties of this 
function, one finds that the hard-scattering kernel satisfies the 
integro-differential equation (for $l_+>0$) \cite{new}\footnote{For 
simplicity of notation, we omit the arguments $E_\gamma$, $m_b$, and 
$x_\gamma$ for the remainder of this section. Also, unless otherwise 
indicated, $\alpha_s\equiv\alpha_s(\mu)$.}
\begin{equation}\label{me}
   \frac{d}{d\ln\mu}\,T(l_+,\mu)
   = \left[ \Gamma_{\rm cusp}(\alpha_s)\,\ln\frac{\mu}{l_+}
   + \gamma(\alpha_s) \right] T(l_+,\mu)
   + \int_0^\infty\!d\omega\,l_+\,\Gamma(\omega,l_+,\alpha_s)\,
   T(\omega,\mu) \,,
\end{equation}
where $\Gamma_{\rm cusp}$ is the universal cusp anomalous dimension 
familiar from the theory of the renormalization of Wilson loops 
\cite{Korchemsky:wg}. The function $\Gamma$ obeys 
$\int d\omega\,\Gamma(\omega,\omega',\alpha_s)=0$. At one-loop order, it 
is given by
\begin{equation}\label{G1l}
   \Gamma^{\rm 1-loop}(\omega,\omega',\alpha_s)
   = - \Gamma_{\rm cusp}^{\rm 1-loop}(\alpha_s)
   \left[ \frac{\theta(\omega-\omega')}{\omega(\omega-\omega')}
   + \frac{\theta(\omega'-\omega)}{\omega'(\omega'-\omega)} \right]_+ ,
\end{equation}
where the plus distribution is defined such that, when $\Gamma$ is 
integrated with a function $f(\omega)$, one must replace 
$f(\omega)\to f(\omega)-f(\omega')$ under the integral. It is 
straightforward to check that our one-loop result in (\ref{Tres}) is a 
solution to (\ref{me}) at order $\alpha_s$. 

From the factorization property of the hard-scattering kernel exhibited 
in (\ref{fact2}) and the functional forms of the hard and jet functions 
given above, it follows that the hard component and the jet function obey 
the RG equations
\begin{eqnarray}\label{Hevol}
   \frac{d}{d\ln\mu}\,H(\mu)
   &=& \left[ - \Gamma_{\rm cusp}(\alpha_s)\,\ln\frac{\mu}{2E_\gamma}
    + \gamma(\alpha_s) - \gamma'(\alpha_s) \right] H(\mu) \,, \\
   \frac{d}{d\ln\mu}\,J(l_+,\mu)
   &=& \left[ \Gamma_{\rm cusp}(\alpha_s)\,
    \ln\frac{\mu^2}{2E_\gamma\,l_+} + \gamma'(\alpha_s) \right]
    J(l_+,\mu) + \int_0^\infty\!d\omega\,l_+\,
    \Gamma(\omega,l_+,\alpha_s)\,J(\omega,\mu) \,. \nonumber
\end{eqnarray}
The reasoning here is analogous to an argument presented by Korchemsky 
and Sterman in their discussion of the $B\to X_s\gamma$ photon spectrum 
\cite{Korchemsky:1994jb}. The anomalous dimensions $\gamma$ and $\gamma'$ 
do not have a simple geometric interpretation and must be determined by 
explicit calculation. From (\ref{Hres}) and (\ref{Jres}) we find
\begin{equation}
   \gamma(\alpha_s) = -2C_F\,\frac{\alpha_s}{4\pi} + O(\alpha_s^2) \,,
    \qquad
   \gamma'(\alpha_s) = O(\alpha_s^2) \,.
\end{equation}

From the discussion above, it follows that in the intermediate theory the 
amplitude is represented in terms of the time-ordered product of the 
electromagnetic current with a heavy-collinear SCET current operator of 
the type $\bar\chi\,\Gamma_1 h$, where $\chi$ is a hard-collinear quark 
field. The corresponding matching relation for such currents reads 
\cite{Bauer:2000yr}
\begin{equation}
   \bar u\gamma^\mu(1-\gamma_5) b
   \to C_3^{\rm SCET}(\mu)\,\bar\chi\gamma^\mu(1-\gamma_5) h + \dots \,,
\end{equation}
where the ellipses represent terms with different Dirac structure. The
Wilson coefficient $C_3^{\rm SCET}(\mu)$ coincides with our function 
$\widetilde H_1(\mu)$. The first relation in (\ref{Hevol}) then 
determines the exact form of the RG equation obeyed by the Wilson 
coefficients of heavy-collinear currents in SCET, which we have derived 
here for the first time. Based on a comparison of one-loop results 
obtained in SCET with expressions presented in \cite{Korchemsky:1994jb} 
for the photon energy spectrum in inclusive $B\to X_s\gamma$ decays, 
previous authors have conjectured a relation between the coefficient of 
the $\ln(\mu/2E_\gamma)$ term in (\ref{Hevol}) and the cusp anomalous 
dimension \cite{Bauer:2000ew,Bauer:2000yr}; however, the nature and 
origin of this connection beyond one-loop order was left unclear. Note, 
in particular, that the minus sign in front of the logarithm in the RG 
equation for $H(\mu)$ does not allow for a simple interpretation in terms 
of a cusp singularity of a Wilson loop. Our derivation establishes two 
important facts: First, to all orders in perturbation theory only a 
single logarithm of the ratio $\mu/2E_\gamma$ appears in the RG equation. 
Secondly, the coefficient of the logarithm equals minus the cusp 
anomalous dimension. The first observation is a crucial one. The fact 
that no higher powers of logarithms appear in the RG equations cannot be 
inferred from a fixed-order calculation in SCET. It follows from 
remarkable properties of Wilson loops discussed long ago by Korchemsky 
and Radyushkin \cite{Korchemsky:wg}. Without this insight it would be 
impossible to integrate the RG equations. The second observation implies 
that the coefficient of the logarithmic term is known to two-loop order, 
which allows for the resummation of Sudakov logarithms at NLO.

We now discuss the general solution of the evolution equations (\ref{me}) 
and (\ref{Hevol}), which is non-trivial due to the convolution integral 
involving the function $\Gamma(\omega,\omega',\alpha_s)$. Note that the 
kernel $\Gamma$ is not simply a function of the difference 
$(\omega-\omega')$, and so the convolution cannot be turned into a 
product using Fourier transformation. Nevertheless, an exact solution can 
be written down in terms of a new function \cite{new}
\begin{equation}
   \F(a,\alpha_s) = \int d\omega\,\omega'\,\Gamma(\omega,\omega',\alpha_s)
   \left( \frac{\omega}{\omega'} \right)^{-a} .
\end{equation}
At one-loop order we find from (\ref{G1l})
\begin{equation}
   \F^{\rm 1-loop}(a,\alpha_s)
   = \Gamma_{\rm cusp}^{\rm 1-loop}(\alpha_s)\,
   \Big[ \psi(1+a) + \psi(1-a) + 2\gamma_E \Big] \,,
\end{equation}
where $\psi(z)$ is the logarithmic derivative of the Euler 
$\Gamma$-function. We start by solving the first equation in 
(\ref{Hevol}) with the initial condition for $H(\mu_h)$ evaluated at a 
high scale $\mu_h\sim m_b$, for which it does not contain large 
logarithms. We then evolve the function $H(\mu)$ down to an intermediate 
scale $\mu_i\sim\sqrt{m_b\Lambda}$ and multiply it by the result 
$J(l_+,\mu_i)$ for the jet function, which at the intermediate scale is 
free of large logarithms and can be written in the general form 
$J(l_+,\mu_i)\equiv\J[\alpha_s(\mu_i),\ln(2E_\gamma\,l_+/\mu_i^2)]$. 
This determines the kernel $T(l_+,\mu_i)$ at the intermediate scale.
Finally, we solve (\ref{me}) and compute the evolution down to a 
low-energy scale $\mu\sim\mbox{few}\times\Lambda_{\rm QCD}$. The exact 
solution is given by
\begin{equation}\label{magic1}
   T(l_+,\mu) = H(\mu_h)\,\J[\alpha_s(\mu_i),\nabla_\eta]\,
   \exp U(l_+,\mu,\mu_i,\mu_h,\eta) \Big|_{\eta=0} \,,
\end{equation}
where the notation $\J[\alpha_s(\mu_i),\nabla_\eta]$ means that one must 
replace each logarithm of the ratio $2E_\gamma\,l_+/\mu_i^2$ by a 
derivative with respect to an auxiliary parameter $\eta$. The evolution 
function $U$ is given by
\begin{eqnarray}\label{magic2}
   U(l_+,\mu,\mu_i,\mu_h,\eta)
   &=& \int\limits_{\alpha_s(\mu_h)}^{\alpha_s(\mu_i)}\!d\alpha\,
    \frac{\Gamma_{\rm cusp}(\alpha)}{\beta(\alpha)}
    \Bigg[ \ln\frac{2E_\gamma}{\mu_h}
     - \int\limits_{\alpha_s(\mu_h)}^\alpha
     \frac{d\alpha'}{\beta(\alpha')} \Bigg]
    - \int\limits_{\alpha_s(\mu_h)}^{\alpha_s(\mu_i)}\!d\alpha\,
    \frac{\gamma'(\alpha)}{\beta(\alpha)} \nonumber\\
   &&\mbox{}- \int\limits_{\alpha_s(\mu_i)}^{\alpha_s(\mu)}\!d\alpha\,
    \frac{\Gamma_{\rm cusp}(\alpha)}{\beta(\alpha)}
    \Bigg[ \ln\frac{l_+}{\mu}
     + \int\limits_\alpha^{\alpha_s(\mu)}
     \frac{d\alpha'}{\beta(\alpha')} \Bigg]
    + \int\limits_{\alpha_s(\mu_h)}^{\alpha_s(\mu)}\!d\alpha\,
    \frac{\gamma(\alpha)}{\beta(\alpha)} \nonumber\\
   &&\mbox{}+ \eta\ln\frac{2E_\gamma\,l_+}{\mu_i^2}
    + \int\limits_{\alpha_s(\mu_i)}^{\alpha_s(\mu)}
    \frac{d\alpha}{\beta(\alpha)}\,
    \F\bigg(-\eta+\!\int\limits_{\alpha_s(\mu_i)}^\alpha\!
    d\alpha'\,\frac{\Gamma_{\rm cusp}(\alpha')}{\beta(\alpha')},
    \alpha\bigg) \,, \qquad
\end{eqnarray}
where $\beta(\alpha_s)=d\alpha_s/d\ln\mu$. Close inspection shows that 
the result for the hard-scattering kernel is independent of the two 
matching scales $\mu_h$ and $\mu_i$. 

Given this exact result, it is straightforward to derive approximate
expressions for the kernel at given orders in RG-improved perturbation 
theory, by using perturbative expansions of the anomalous dimensions and 
$\beta$-function to the required order. Unfortunately, controlling terms 
of $O(\alpha_s)$ in the evolution function $U$ would require knowledge 
of the cusp anomalous dimension at three-loop order (as well as 
knowledge of all other anomalous dimensions at two-loop order), which at 
present is lacking. We will, however, control the dependence on the
variables $l_+$ and $E_\gamma$ to $O(\alpha_s)$. As usual, we write
\begin{equation}
   \beta(\alpha_s) = -2\alpha_s \sum_{n=0}^\infty \beta_n
    \left( \frac{\alpha_s}{4\pi} \right)^{n+1} , \qquad
   \Gamma_{\rm cusp}(\alpha_s) = \sum_{n=0}^\infty \Gamma_n
    \left( \frac{\alpha_s}{4\pi} \right)^{n+1} ,
\end{equation}
and similarly for the anomalous dimensions $\gamma$ and $\gamma'$. The
relevant expansion coefficients are $\Gamma_0=4C_F$, $\Gamma_1=4C_F
\big[(\frac{67}{9}-\frac{\pi^2}{3})\,C_A-\frac{20}{9}\,T_F n_f\big]$, 
$\gamma_0=-2C_F$, $\gamma_0'=0$, and 
$\beta_0=\frac{11}{3}\,C_A-\frac43\,T_F n_f$, 
$\beta_1=\frac{34}{3}\,C_A^2-\frac{20}{3}\,C_A T_F n_f-4C_F T_F n_f$.
Defining the ratios $r_1=\alpha_s(\mu_i)/\alpha_s(\mu_h)$ and
$r_2=\alpha_s(\mu)/\alpha_s(\mu_i)$, we obtain our final result
\begin{eqnarray}\label{exact1}
   T(l_+,\mu)
   &=& e^{U_0(\mu,\mu_i,\mu_h)} \left( \frac{l_+}{\mu} \right)^{c\ln r_2}
    \left( \frac{2E_\gamma}{\mu_h} \right)^{-c\ln r_1} \nonumber\\
   &\times& \Bigg\{ 1 + \frac{C_F\,\alpha_s(\mu_h)}{4\pi}
    \bigg[ - 2\ln^2\frac{2E_\gamma}{\mu_h} + 2\ln\frac{2E_\gamma}{\mu_h} 
    - \frac{x_\gamma\ln x_\gamma}{1-x_\gamma} - 2L_2(1-x_\gamma)
    - 4 - \frac{\pi^2}{12} \bigg] \nonumber\\
   &&\hspace{0.6cm}\mbox{}+ \frac{C_F\,\alpha_s(\mu_i)}{4\pi}
    \bigg[ \left( \ln\frac{2E_\gamma\,l_+}{\mu_i^2}
    - \psi(1+c\ln r_2) - \psi(1-c\ln r_2) - 2\gamma_E \right)^2
    \nonumber\\
   &&\hspace{3.1cm}\mbox{}- \psi'(1+c\ln r_2) + \psi'(1-c\ln r_2)
    - 1 - \frac{\pi^2}{6} \bigg] \\
   &&\hspace{0.6cm}\mbox{}+ \frac{\Gamma_0}{2\beta_0}
    \left( \frac{\Gamma_1}{\Gamma_0} - \frac{\beta_1}{\beta_0} \right)
    \bigg[ \frac{\alpha_s(\mu)-\alpha_s(\mu_i)}{4\pi}\,\ln\frac{l_+}{\mu}
    - \frac{\alpha_s(\mu_i)-\alpha_s(\mu_h)}{4\pi}\,
    \ln\frac{2E_\gamma}{\mu_h} \bigg] \Bigg\} \,, \nonumber
\end{eqnarray}
where $c=\Gamma_0/2\beta_0$, and 
\begin{eqnarray}\label{exact2}
   U_0(\mu,\mu_i,\mu_h) 
   &=& \frac{\Gamma_0}{4\beta_0^2}\,\Bigg\{
    (1-\ln r_1)\,\frac{4\pi}{\alpha_s(\mu_h)}
    + (1+\ln r_2)\,\frac{4\pi}{\alpha_s(\mu)}
    - \frac{8\pi}{\alpha_s(\mu_i)} \nonumber\\
   &&\hspace{1.0cm}\mbox{}+
    \frac{\beta_1}{2\beta_0}\,(\ln^2 r_1+\ln^2 r_2)
    + \left( \frac{\Gamma_1}{\Gamma_0} - \frac{\beta_1}{\beta_0}
    \right) \left( \ln\frac{r_1}{r_2} + 2 - r_1 - \frac{1}{r_2} \right)
    \Bigg\} \nonumber\\
   &&\mbox{}- \frac{\gamma_0}{2\beta_0}\,\ln(r_1 r_2) 
    - \ln\frac{\Gamma(1+c\ln r_2)}{\Gamma(1-c\ln r_2)} 
    - 2\gamma_E\,c\ln r_2 + O(\alpha_s)
\end{eqnarray}
corresponds to the evolution function $U$ evaluated with $\eta=0$, 
$l_+=\mu$, and $2E_\gamma=\mu_h$. The only piece missing for a complete 
resummation at NNLO is the $O(\alpha_s)$ contribution to $U_0$, which is 
independent of $l_+$ and $E_\gamma$.

\begin{figure}
\epsfxsize=15.0cm
\centerline{\epsffile{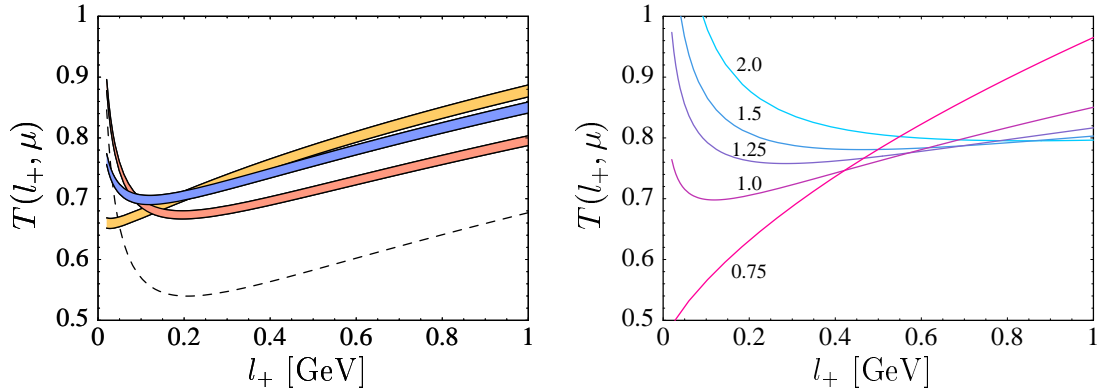}}
\vspace{0.0cm}
\centerline{\parbox{14cm}{\caption{\label{fig:resum}
RG-improved predictions for the hard-scattering kernel at maximum photon
energy. 
{\em Left:\/} Results at $\mu=1$\,GeV. The bands refer to different 
values of the intermediate matching scale: $\mu_i^2=\Lambda_h m_b$ 
(center), $2\Lambda_h m_b$ (top), $0.5\Lambda_h m_b$ (bottom). Their 
width reflects the sensitivity to the high-energy matching scale 
$\mu_h^2$, varied between $2m_b^2$ and $0.5m_b^2$. The dashed line shows 
the result obtained at one-loop order. 
{\em Right:\/} Dependence of the kernel on the renormalization scale 
$\mu$, varied between 0.75\,GeV and 2.0\,GeV as indicated on the 
curves.}}}
\end{figure}

In order to study the importance of RG improvement and Sudakov 
resummation, we compare in the left-hand plot in Figure~\ref{fig:resum} 
the result for the resummed hard-scattering kernel in (\ref{exact1}) with 
the one-loop approximation in (\ref{Tres}). We choose the highest
possible value 
of the photon energy ($E_\gamma=m_b/2$ with $m_b=4.8$\,GeV) so as to 
maximize the values of the large logarithms, and plot the function 
$T(l_+,\mu)$ for $\mu=1$\,GeV and different choices of the matching 
scales $\mu_h$ and $\mu_i$. Here and below, the ``natural'' choices 
$\mu_h=2E_\gamma$ and $\mu_i=\sqrt{2E_\gamma\Lambda_h}$, where 
$\Lambda_h=0.5$\,GeV serves as a typical hadronic scale, are taken as 
default values. We use the two-loop running coupling normalized at 
$\alpha_s(m_b)=0.22$ and set $n_f=4$ for the number of light quark 
flavors. (For simplicity, we do not match onto a three-flavor theory even 
for low renormalization scales.) We find that resummation effects 
decrease the magnitude of the radiative corrections, i.e., the resummed 
kernel is closer to the tree-level value ($T=1$) than the one-loop 
result. The fact that after Sudakov resummation the radiative corrections 
are moderate in magnitude persists even for asymptotically large 
$b$-quark masses. For instance, setting $m_b=50$\,GeV we find after 
resummation $T(l_+,\mu)=0.74$ at $l_+=\mu=1$\,GeV. (Fixed-order 
perturbation theory breaks down for such large values of the quark mass. 
From (\ref{Tres}) we would obtain $T(l_+,\mu)=0.08$ with these parameter 
values.) The figure also exhibits that our results are stable under 
variation of the two matching scales. Varying $\mu_i^2$ and $\mu_h^2$ by 
factors of 2 changes the result for the kernel by less than 10\%. This 
suggests that the unknown NNLO corrections to the function $U_0$ in 
(\ref{exact2}) are perhaps not very important.

The scale dependence of the resummed expression for the kernel is 
illustrated in the right-hand plot in Figure~\ref{fig:resum}, which shows 
the functional dependence of $T(l_+,\mu)$ for maximal photon energy and 
several values of $\mu$. The matching scales are set to their default 
values $\mu_h=m_b=4.8$\,GeV and 
$\mu_i=\sqrt{\Lambda_h m_b}\simeq 1.55$\,GeV. We observe a significant 
scale dependence of the kernel, especially as one lowers $\mu$ below the 
intermediate scale $\mu_i$. In other words, the second stage of running 
(for $\mu<\mu_i$), which we have computed for the first time in the 
present paper, is numerically significant. 

\begin{figure}
\epsfxsize=8.0cm
\centerline{\epsffile{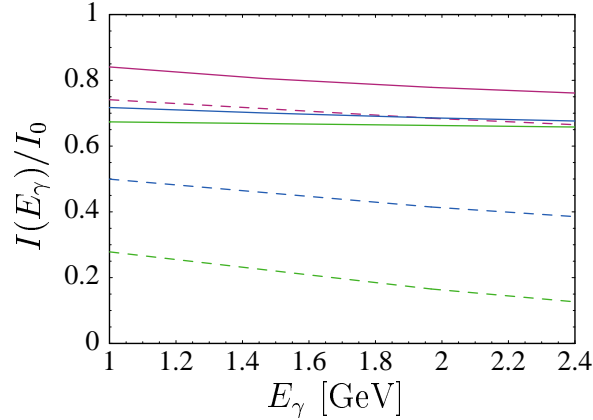}}
\vspace{0.0cm}
\centerline{\parbox{14cm}{\caption{\label{fig:lamB}
Energy dependence of the convolution integral $I(E_\gamma)$ normalized
to its tree-level value, assuming the model (\ref{model}) for the LCDA
at a low scale $\mu_0$ such that $\alpha_s(\mu_0)=0.5$ (top), 0.75 
(center), and 1.0 (bottom). The matching scales are set to their default 
values. The solid curves correspond to the resummed kernel, while the 
dashed ones are obtained at one-loop order.}}}
\end{figure}

Finally, it is interesting to study resummation effects for the 
convolution integral
\begin{equation}
   I(E_\gamma) = \int_0^\infty\!dl_+\,\frac{\phi_+^B(l_+,\mu)}{l_+}\,
   T(l_+,E_\gamma,m_b,\mu) \,. 
\end{equation}
At tree-level, this integral has been denoted by $I_0=1/\lambda_B$, where 
$\lambda_B$ is a low-energy hadronic parameter \cite{Beneke:1999br}. To 
evaluate the integral beyond tree-level one needs to assume a particular 
form of the LCDA. In \cite{Grozin:1996pq}, a model function was derived 
from a QCD sum-rule analysis of the matrix element of the bilocal HQET 
operator in (\ref{LCDA}). This study motivated the ansatz
\begin{equation}\label{model}
   \phi_+^B(l_+,\mu_0) = \frac{l_+}{\lambda_B^2}\,e^{-l_+/\lambda_B} \,,
   \quad \mbox{with} \quad
   \lambda_B = \frac23\,(m_B-m_b)\approx 0.32\,\mbox{GeV} \,.
\end{equation}
Such a nonperturbative calculation, which does not control the scale 
dependence of the LCDA, is reasonable only at a low hadronic scale 
$\mu_0$. For larger values of $\mu$ evolution effects introduce a
radiative tail in the LCDA. We thus expect significant modifications 
of the convolution integral due to radiative corrections. While the
function $I(E_\gamma)$ is formally $\mu$-independent, our model estimate 
will depend on the value of the scale $\mu_0$ at which the functional 
form given in (\ref{model}) is assumed to be correct. In 
Figure~\ref{fig:lamB}, we show results for the function $I(E_\gamma)$ in 
units of $I_0$ for three different values of $\mu_0$. After RG 
resummation we observe a modest reduction of $I(E_\gamma)$ with respect 
to its tree-level value, which is fairly insensitive to the precise value 
of $\mu_0$ and only shows a mild energy dependence. In a rough 
approximation we have $I(E_\gamma)\approx 0.75 I_0$.
In contrast, the results obtained at one-loop order are strongly 
sensitive to the choice of $\mu_0$ and exhibit a more pronounced 
dependence on the photon energy.

\section{Conclusions}

We have applied soft-collinear effective theory to prove a QCD 
factorization formula for the radiative semileptonic decay 
$B\to\gamma l\nu$, stating that at leading power in $\Lambda/m_b$ the 
decay amplitude can be written as a convolution of a perturbative 
hard-scattering kernel with the leading-order $B$-meson light-cone 
distribution amplitude. Besides deriving the precise form of this 
convolution and establishing that the kernel is infrared finite to all 
orders in perturbation theory, we have shown that the convolution 
integral is free of endpoint singularities, and that non-valence Fock 
states of the $B$ meson do not contribute at leading power.

We believe that the analysis of the factorization properties of the decay 
amplitude is most transparent in the framework of the formulation of 
soft-collinear effective theory developed in \cite{Hill:2002vw}, where 
full QCD is directly matched onto a low-energy effective theory in which 
only long-distance modes are kept as dynamical degrees of freedom. In the 
present case these are the soft constituents of the $B$ meson and the 
collinear photon field. For more complicated processes with energetic, 
light final-state mesons one would also have to introduce collinear quark 
and gluon fields. Many of the techniques used in our analysis (such as 
soft-collinear gauge invariance, reparameterization invariance, and
renormalization-group improvement) are equally relevant in such a more 
general context.

The second part of our analysis was devoted to the calculation of the 
hard-scattering kernel in the factorization formula using 
renormalization-group improved perturbation theory. We have established a 
second, perturbative factorization formula, according to which the 
different short-distance scales entering in the calculation of the kernel 
(hard scales of order $m_b$ and hard-collinear scales of order 
$\sqrt{m_b\Lambda}$) can be separated into a hard function and a jet 
function. The corresponding two classes of large logarithms can be 
systematically resummed by solving evolution equations derived from the 
renormalization properties of the leading-order $B$-meson light-cone 
distribution amplitude. As a byproduct, we have elucidated the relation 
between the anomalous dimension of heavy-collinear currents in the 
effective theory and the universal cusp anomalous dimension encountered 
in the study of Wilson loops with light-like segments. In contrast to 
previous analyses, we have performed a complete resummation of Sudakov 
logarithms down to a low-energy scale 
$\mu\sim\mbox{few}\times\Lambda_{\rm QCD}$ independent of the heavy-quark 
mass. Only in that way all dependence on $m_b$ and the photon energy 
$E_\gamma$ is explicitly contained in the hard-scattering kernel. Our 
results (\ref{magic1}) and (\ref{magic2}) give the exact analytic 
solution for the kernel, valid to all orders in perturbation theory.

The discussion of the decay $B\to\gamma l\nu$ presented here can be taken 
over almost verbatim to analyze related processes such as 
$B\to\gamma\gamma$ and $B\to\gamma\,l^+ l^-$. The interesting observation 
that at leading power the long-distance effects in these processes are 
universal (for fixed photon energy) \cite{Descotes-Genon:2002ja} can be 
understood in our formalism as follows: Once short-distance modes related 
to hard and collinear interactions with the heavy quark are integrated 
out at a scale $\mu_h\sim m_b$, the evolution of the resulting operators 
in the effective theory is independent of the Dirac structure of the 
relevant current operators. This is true for the evolution of 
heavy-collinear currents (running from $\mu_h$ down to an intermediate 
scale $\mu_i\sim\sqrt{m_b\Lambda}$), and for the bilocal heavy-light 
currents whose matrix elements are expressed in terms of the 
leading-order $B$-meson light-cone distribution amplitude (evolution from 
$\mu_i$ down to a low-energy hadronic scale). Process-dependent 
corrections arise only in the initial matching at the high-energy scale 
and thus are calculable in an expansion in powers of $\alpha_s(m_b)$. 

From a phenomenological point of view, an important finding of our 
analysis is that Sudakov resummation does not lead to a strong 
suppression of the decay amplitudes. After renormalization-group 
improvement we find moderate corrections to the kernel which are smaller 
than at one-loop order, and which stay at the level of 20--30\% even in 
the hypothetical case of asymptotically large heavy-quark mass.

\vspace{0.3cm}  
{\it Acknowledgment:\/} 
We are grateful to Thomas Becher for useful discussions. The research of 
S.W.B., B.O.L.\ and M.N.\ was supported by the National Science 
Foundation under Grant PHY-0098631. The research of R.J.H.\ was supported 
by the Department of Energy under Grant DE-AC03-76SF00515.

\end{document}